\documentclass[12pt,eps]{article}
\usepackage{authblk}

\begin{document}

\setcounter{page}{1}

\title{On a Broken Formal Symmetry between Kinetic and Gravitational Energy\footnote{This is the revised version of an essay originally written for the 2010 Gravity Research Foundation Essay Competition}}

\author{Armin Nikkhah Shirazi\thanks{electronic mail: armin@umich.edu }}

\affil{\small \textit{Department of Physics}\\ \textit{University of Michigan} \\  \textit{450 Church Street}\\ \textit{ Ann Arbor MI 48109}}

\maketitle
\small \date{}

\normalsize

\begin{abstract}
Historically, the discovery of symmetries has played an important role in the progress of our fundamental understanding of nature. This paper will demonstrate that one can construct in Newtonian theory in a spherical gravitational field a formal symmetry between the kinetic (KE) and gravitational potential energy (GPE) of a test mass.  Put differently, one can construct a way of expressing GPE such that the \emph{form of the mathematical expression} remains invariant under an interchange of KE and GPE. When extended to relativity by a suitable axiom, it leads to a framework that bridges the general relativistic and Newtonian conceptions of gravitational energy, even though the symmetry is broken except in the infinitesimal limit. Recognizing this symmetry at infinitesimal scales makes it possible to write the gravitational energy-momentum relation which presumably pertains to the graviton, the properties of which under one interpretation may be unexpected.
\end{abstract}
\section{Introduction}

Historically, the discovery of symmetries has played an important role in the progress of our fundamental understanding of nature \cite{Zee07}, but generally these symmetries, and their breaking, are intrinsic aspects of the structure of particular theories \cite{And72}, and so one could call them \emph{structural symmetries}. In contrast, this paper will demonstrate that one can construct in Newtonian theory in a spherical gravitational field a \emph{formal symmetry} between the kinetic (KE) and gravitational potential energy (GPE) of a test mass.  Put differently, one can construct a way of expressing GPE such that the \emph{form of the mathematical expression} remains invariant under an interchange of KE and GPE, though structurally they are still completely different concepts.  When extended to relativity by a suitable axiom, the formal symmetry leads to a framework that links the general relativistic and Newtonian conceptions of gravitational energy in a novel way, even though the symmetry is broken except in the infinitesimal limit. Recognizing this symmetry at infinitesimal scales makes it possible to write the gravitational energy-momentum relation which presumably pertains to the graviton, the properties of which under one interpretation may be unexpected.

\section{The Formal Symmetry in Newtonian Physics}

We assume that the gravitational force is mediated by the gravitational field. Let us define the force exerted by the field on the test mass by $\mathbf{F}$ and that exerted by the test mass on the field by $\mathbf{F^*_g}$. Then, Newton's third law says
\begin{equation}
\mathbf{F=-F^*_g}
\end{equation}

We also know that \textbf{F} is the negative gradient of the GPE
\begin{equation}
\mathbf{F}=\mathbf{-\nabla}U=-m\nabla\phi
\end{equation}

Where $U$ is the GPE and $\phi$ is the gravitational potential. In  Newtonian theory there exists no finite limit on motion, so fields as mediators are little more than mathematical artifacts: one simply has \emph{action at a distance} which due to the infinite transmission speed renders the field concept physically hollow.  Moreover, in Newtonian theory GPE is usually associated with boundary conditions which require it to be negative whereas mass is a positive quantity, and this introduces certain challenges in constructing in a mathematically consistent fashion expressions for gravitational quantities that are formally symmetric to kinetic energy. 
\\To address these challenge, and for reasons of mathematical consistency, the momentum stored in this field, $\mathbf{p_g}$, must be considered an imaginary quantity, so that to relate $\mathbf{F^*_g}$ to $\mathbf{p_g}$, we need to define an imaginary force term $\mathbf{F_g}$ such that
\begin{equation}
\mathbf{F^*_g}=-i \mathbf{F_g}=-i \frac{d\mathbf{p_g}}{dt}=m\nabla\phi
\end{equation}

Where we used (2) to relate the gradient of the field's potential Energy to the momentum stored in it, and inserted the factor $-i$ to ensure that $\mathbf{F^*_g}$ is real. The interpretation of $\mathbf{F_g}$ is that it is more intrinsically tied to the mediating field, which in Newtonian theory is little more than a contrivance, after all. This seems to be the price to pay for superimposing a mediator onto a theory which has no need for it but it is indispensable for constructing the formal symmetry.
\\Let us now assume the simplest field configuration, \emph{spherical symmetry}. Using the chain rule on the left under the justification that the momentum stored in the field can be thought of as function of the position of the test mass in the field, and writing the gradient in terms of a change in the radial direction $\mathbf{\widehat{r}}$ only  gives
\begin{equation}
\frac{d \mathbf{p_g}}{dt}=\frac{dr}{dt}\frac{d\mathbf{ p_g}}{dr}=i \hspace{0.01in}m\frac{d\phi}{dr}\widehat{\mathbf{r}}
\end{equation}

Multiplying both sides by $mdr$ and defining $m\frac{dr}{dt}\equiv{mv_r}\equiv{p_r}$  gives
\begin{equation}
p_rd\mathbf{p_g}=i \hspace{0.01in}m^2d\phi\widehat{\mathbf{r}}
\end{equation}

We wish to integrate this in such a way as to obtain an expression for GPE purely in terms of $p_g$. To do so, first recall that (1) applied to this situation can be rewritten as
\begin{equation}
\frac{d\mathbf{p_r}}{dt}=-\left ( -i\frac{d\mathbf{p_g}}{dt} \right )=i\frac{d\mathbf{p_g}}{dt} 
\end{equation}

Or, considering just the differential momenta,
\begin{equation}
d\mathbf{p_r}=i \hspace{0.01in}d\mathbf{p_g}
\end{equation}

Which upon integration yields
\begin{equation}
\mathbf{p_r}=i \hspace{0.01in}\mathbf{p_g}+p_0
\end{equation}

Where $p_0$ is an integration constant with dimensional units of momentum.  The interpretation of this constant depends on the boundary conditions we impose. We assume the usual BCs, namely that at infinity the test mass is at rest and that the potential is set to zero, which means the stored momentum of the field there is also set to zero. Since due to our BCs there exists a region where both momentum variables are zero, $p_0$ must vanish, leaving
\begin{equation}
\mathbf{p_r}=i \hspace{0.01in}\mathbf{p_g}
\end{equation}

This implies
\begin{equation}
p_r=i \hspace{0.01in}p_g
\end{equation}

Substituting (10) into (5) and integrating both sides in the radial direction from zero to infinity then gives
\begin{equation}
\frac{p^2_g}{2}=m^2\phi
\end{equation}

where the integration constant is again suppressed by our previous BCs. Dividing through by $m^2$ gives
\begin{equation}
\frac{p^2_g}{2m^2}\equiv{\frac{v^2_g}{2}}=\phi
\end{equation}

where 
\begin{equation}
v_g\equiv{\frac{p_g}{m}}
\end{equation}

is the stored momentum per test mass $m$ and will be defined as \emph{the motion stored in the gravitational field}. It can in the Newtonian context be thought of as the `potential motion' of the test mass i.e. motion which is `gained' by the field as the test mass moves from the source to infinity. Notice that the consideration of  $v_g$ as an imaginary quantity is consistent with the fact that $\phi<0$. We will show at the end of the next section that it is possible to think of this as a mathematical artifact.  For a potential given by Newton's law of gravitation
\begin{equation}
v_g=\sqrt{2\phi}=\sqrt{\frac{-2GM}{r}}
\end{equation}

By defining $v_g$ we can write the total test mass energy in Newtonian theory as
\begin{equation}
E=K+U=\frac{1}{2}mv^2_r-G\frac{Mm}r=\frac{1}{2}mv^2_r+\frac{1}{2}mv^2_g
\end{equation}

making the formal symmetry between classical KE and GPE for the spherical field explicit. Due to our BCs $E=0$. For different BCs, the value of $E$ might differ due to non-vanishing integration constants, but the symmetry remains as long as $E$ is constant, which is just a statement  of conservation of energy.

\section{The Formal Symmetry in Relativity}

To develop the same idea in relativity we assume that the formal symmetry as shown in (15) has its origin in a corresponding formal symmetry in relativistic momentum. Formally, we assume \emph{for point-like particles only}
\begin{equation}
\textbf{Axiom:   }\\\gamma_gmv_g\mbox{ (Relativity)}\rightarrow{}mv_g\mbox{(Newtonian)}
\end{equation}

where we take $v_g$ in relativity to be real and
\begin{equation}
\gamma_g\equiv{\frac{1}{\sqrt{1-\frac{v^2_g}{c^2}}}}=\frac{dt}{d\tau(K=0,U\neq{0})}
\end{equation}

Is the gravitational analog to the Lorentz factor $\gamma$: for $K=0,U=0,\gamma_g\rightarrow{1}$ and for  $K>0,U=0,\gamma_g\rightarrow{\gamma}$ where $U$, the gravitational energy in this context, will shortly be identified with $U_+$ in equation $(20)$, just as $K$ is identified with one of two possible solutions of a quadratic equation. To preserve the formal symmetry, we set $K=0$,  throughout. In a rest frame in a gravitational field, by formal symmetry with kinetic energy, we write
\begin{equation}
E^2=(mc^2+U)^2=m^2c^4+p^2_gc^2
\end{equation}

We emphasize that by treating gravitational energy as a property of the test mass, we are adopting a perspective radically different from that of canonical GR, which views it as a property of the spacetime region in which the mass finds itself. We adopt this perspective to explore this formal symmetry in relativity. To obtain the relativistic expression for $U$, rewrite (18) as
\begin{equation}
U^2+2Umc^2-p^2_gc^2=0
\end{equation}

This quadratic equation in $U$ has two distinct roots. The first is
\begin{equation}
U_+=mc^2(\gamma_g-1)
\end{equation}

This solution is exactly symmetric in form to relativistic kinetic energy. It is non-negative and decreases with increasing distance to zero infinitely far from the gravitational source. Furthermore, given that
\begin{equation}
ds^2(K=0,U>0)=\frac{c^2dt^2}{\gamma_g^2}
\end{equation}

where $ds^2$ is the spacetime interval, we should expect changes in  $U_+$ to be associated with  a change in the geometry of spacetime in a manner that is qualitatively consistent with canonical GR, as for instance exemplified by gravitational time dilation.  Even though quantifying the change requires a re-expression of $U_+$ in terms of Energy density, this solution seems very close in spirit to the general relativistic conception of gravitational Energy. 
\\The second solution is
 \begin{equation}
U_-=-mc^2(\gamma_g+1)
\end{equation}

This solution differs from (20) by a sign in $\gamma_g$. It cannot be associated with the same geometric interpretation as  $U_+$ because it is negative and increases with increasing distance. These are characteristics of Newtonian GPE. Notice that infinitely far from the source, where  the field is zero, $U_-$ increases to a maximum value of  $-2mc^2$, so that in a zero field the total energy of the test mass becomes $E=mc^2+(-2mc^2)=-mc^2$. But since the total energy here describes  \emph{a mass at rest in a zero field}, this is equivalent to considering the `net' rest mass to be negative, i.e. the mass associated with rest energy $E$ is $-m$ if one chooses the second solution.  Expanding $(22)$ to first order we get
\begin{equation}
U_-\approx -m \left ( \frac{1}{2}{v_g}^2+2c^2 \right )
\end{equation}
 
Where the second term on the right can be cancelled by the arbitrary additive constant $2mc^2$ (Indeed the fact that one way to characterize the transition from relativity to classical physics is to take $c \rightarrow \infty$ further underlines the arbitrariness of such a constant in this context). In the absence of such constants, both sides are negative: The left side because $U_-$ is always negative except when the test mass is at infinity, where $v_g=0$, and the right side because the choice of $U_-$ for the gravitational energy requires the `net' rest mass to be negative. But in the Newtonian approximation we consider mass to be a positive quantity, and replacing $-m$ by $m$ while keeping the same interpretation of $U$-namely as $U_-$ as opposed to as $U_+$, the first order expansion of which is in fact just $\frac{1}{2}mv^2_g$-forces us to interpret $v_g$ to be imaginary in order for the sign of both sides to match in this approximation. Hence, the imaginary appearance of $v_g$ in Newtonian theory can be thought of as an artifact brought about by a mismatch in the signs of mass and gravitational energy, as mentioned above. Let us  however briefly point out that it \emph{is} possible to interpret $v_g$ as an imaginary quantity even in a relativistic context under the same axiom as above. In that case, we must replace $\gamma_g$ by

\begin{equation}
{\gamma^*_g}\equiv{\frac{1}{\sqrt{1+\frac{v^2_g}{c^2}}}}
\end{equation}
and equation $(18)$ by
\begin{equation}
E^2=(mc^2+U)^2=m^2c^4-{p^*_g}^2 c^2
\end{equation}
Since now ${p^*_g}=\gamma^*_g mv_g$ is imaginary. The solutions, re-expressed under this approach, are
\begin{equation}
U_+=mc^2({\gamma_g}^*-1)
\end{equation}

and 
 \begin{equation}
U_-=-mc^2({\gamma_g}^*+1)
\end{equation}

However, because this approach requires gravitational momenta to be imaginary, it  seems less palatable (especially in light of the next section) and we will therefore continue to consider $v_g$ to be real in relativity.
\\As a final check, use the fact known from canonical GR that for a static spherical field
\begin{equation}
\gamma_g=\frac{1}{\sqrt{1-\frac{2GM}{rc^2}}}
\end{equation}

And take the negative gradient of (22) using (28):
\begin{equation}
-\nabla(-(mc^2(\gamma_g+1)))=-G\frac{Mm}{r^2(1-\frac{2GM}{rc^2})^\frac{3}{2}}\widehat{\mathbf{r}}=-\gamma^3_gG\frac{Mm}{r^2}\widehat{\mathbf{r}}
\end{equation}

for a weak field $(\gamma_g \approx 1)$this reduces to Newton's law of gravitation. The gradient operator is non-relativistic because it communicates changes in the potential instantaneously, but the time-independence of $\gamma_g$ mitigates the importance of this distinction. 
Equation $(16)$  thus leads to a framework that conceptually links the seemingly very different notions of gravitational Energy in Newtonian theory and General Relativity.

Alas, in relativity the formal symmetry between $K$ and $U$ is broken at all but infinitesimal scales:  Viewed as a consequence of our axiom, the symmetry can only hold when the total energy of a test mass divided by its volume describes its energy density throughout, so that the description of  its energy and momentum is essentially linear.  The test mass must therefore be infinitesimally small. If it has finite extent, the axiom given by $(16)$ becomes inapplicable as the associated energies arise from non-linear energy density distributions, and one must resort to a description in terms of the energy-momentum tensor instead, as prescribed by canonical GR \cite{Mis73}. In Newtonian theory this is not a problem because test masses are taken to be point-like from the outset.
 
\section{The Graviton Energy-Momentum Relation}

Given the axiom in equation $(16)$, in the infinitesimal limit and for $m=0$ in analogy to the photon equation, $(18)$ reduces to
\begin{equation}
E=p_gc
\end{equation}

Like a photon, the object presumably described by this equation has zero inertial mass but unlike a photon its momentum and energy are purely gravitational. It is therefore natural to identify this as a graviton, a particle of gravitation. 
This equation is not evident from within standard GR because GR views gravitational energy as a property not of a test mass but of spacetime. Also, GR is most applicable when the formal symmetry is broken. 

At first glance, the form of $(30)$ suggests a particle which travels in space at the speed of light, consistent with our current ideas about the properties of gravitons. That requires, however, that in the limit of $c$, $p_g\rightarrow{p}$, which means that the graviton's momentum and energy are then \emph{kinetic}. At least in this author's view, this presents a conceptual problem because by its very definition one \emph{should} expect the graviton's energy to be gravitational and not kinetic. 
\\There is a second possible interpretation which avoids this problem. By this second possibility, $p_g$ is interpreted in the limit of $c$ the same as it is within the context of massive particles. In that case, the graviton's momentum describes its position, which is at a horizon.  This appears to be the case in general since for $v_g=c$, $\gamma_g$ is generically singular. While the singularities associated with horizons can be transformed away, the underlying  spacetime features which gave rise to them under the choice of certain coordinate systems cannot, and the second interpretation directly associates these features with the presence of gravitons. 
How does the graviton mediate the field under the second interpretation? Classically, at least, the mechanism is straightforward: Since $ds^2=0$ corresponds to a spherical surface expanding at the speed of light, given enough time the graviton's presence is eventually communicated to any region in space by virtue of the geometric structure of spacetime, i.e.\emph{ by virtue of the spatially outward propagating change in the structure of spacetime alone,} without requiring it to have kinetic energy! 
\\It is clear that the second interpretation would force a significant reassessment of some of our current ideas about quantum gravity \cite{Woo09}. However, the same interpretation may also hold the key to understanding another deep problem in canonical GR: It has long been understood that the horizons of Black holes are associated with entropy \cite{Bek73} and with temperature \cite{Haw74}\cite{Haw75}. While various independent approaches have confirmed this \cite{Car07}, an open question that remains is what the microscopic degrees of freedom are which give rise to these thermodynamic phenomena \cite{Pad08}. The object described by equation $(30)$ would seem under the second interpretation to be a nearly ideal candidate, as it is necessarily both pointlike and always located at a horizon. Identifying the state of this object with a thermodynamic black hole microstate, however, may well require an extension if not outright revision of certain thermodynamic concepts as they pertain to black holes. For example, it seems almost inevitable that taking this possibility seriously would require the concept of temperature to be extended to one that permits it to be considered a function of average motion stored in the gravitational field. 
\\In summary, while unexpected, the second interpretation of the object described by equation (30) seems to offer the prospect of exciting novel approaches to understanding quantum gravity and black hole thermodynamics more deeply.

\section{Conclusion}

This paper demonstrated that under the assumption of a formal symmetry between relativistic kinetic and gravitational momentum one can construct a formal symmetry between $K$ and $U$ which helps clarify the relation between the general relativistic and Newtonian conceptions of gravitational Energy, yielding for spherical fields the Newtonian description as an approximation, and which, while broken at non-infinitesimal scales, yields a gravitational Energy-momentum relation that can be interpreted to characterize the graviton. As in the photon's case, the relation does not admit definite energy and momentum values, but it offers the prospect of exciting novel approaches to understanding quantum gravity and black hole thermodynamics more deeply.

\end{document}